\documentclass[twocolumn,letterpaper,amsmath,amssymb,floatfix,aps,superscriptaddress]{revtex4}
\pdfoutput=1
\usepackage{graphicx}
\usepackage{dcolumn}
\usepackage{bm}
\usepackage{epsfig,color}

\def\rmd{{\mathrm{d}}}

\def\rme{{\mathrm{e}}}

\def\Eq{Eq.}
\def\Eqs{Eqs.}
\def\Figure{Figure}
\def\Fig{Fig.}

\def\SItext{{\em Supporting Information}}
\def\ie{{\em i.e.}}
\def\eg{{\em e.g.}}

\newcommand{\kB}{k_\mathrm{B}}

\newcommand{\trm}[1]{{\textrm{#1}}}

\usepackage{soul}  

\begin{document}

\title{Selective solute adsorption and partitioning around single PNIPAM chains}

\author{\rm Matej Kandu\v c}
\affiliation{\rm\small Soft Matter and Functional Materials, Helmholtz-Zentrum Berlin f\"ur Materialien und Energie, Hahn-Meitner-Platz 1, D-14109 Berlin, Germany}
\email{matej.kanduc@helmholtz-berlin.de; joachim.dzubiella@helmholtz-berlin.de}

\author{Richard Chudoba}
\affiliation{\rm\small Soft Matter and Functional Materials, Helmholtz-Zentrum Berlin f\"ur Materialien und Energie, Hahn-Meitner-Platz 1, D-14109 Berlin, Germany}
\affiliation{\rm\small Institut f\"ur Physik, Humboldt-Universit\"{a}t zu Berlin, Newtonstr.\ 15, D-12489 Berlin, Germany}

\author{Karol Palczynski}
\affiliation{\rm\small Soft Matter and Functional Materials, Helmholtz-Zentrum Berlin f\"ur Materialien und Energie, Hahn-Meitner-Platz 1, D-14109 Berlin, Germany}

\author{Won Kyu Kim}
\affiliation{\rm\small Soft Matter and Functional Materials, Helmholtz-Zentrum Berlin f\"ur Materialien und Energie, Hahn-Meitner-Platz 1, D-14109 Berlin, Germany}

\author{Rafael Roa}
\affiliation{\rm\small Soft Matter and Functional Materials, Helmholtz-Zentrum Berlin f\"ur Materialien und Energie, Hahn-Meitner-Platz 1, D-14109 Berlin, Germany}

\author{Joachim Dzubiella}
\affiliation{\rm\small Soft Matter and Functional Materials, Helmholtz-Zentrum Berlin f\"ur Materialien und Energie, Hahn-Meitner-Platz 1, D-14109 Berlin, Germany}
\affiliation{\rm\small  Institut f\"ur Physik, Humboldt-Universit\"{a}t zu Berlin, Newtonstr.\ 15, D-12489 Berlin, Germany}

\begin{abstract}
Thermoresponsive polymer architectures have become integral building blocks of `smart' functional materials in modern applications.  For a large range of developments, \eg, for drug delivery or nanocatalytic carrier systems, the selective adsorption and partitioning of molecules (ligands or reactants) inside the polymeric matrix are key processes that have to be controlled and tuned for the desired material function. In order to gain insights into the nanoscale structure and binding details in such systems, we here employ molecular dynamics simulations of the popular poly(N-isopropylacrylamide) (PNIPAM) polymer in explicit water in the presence of various representative solute types with focus on aromatic model reactants. 
We model a PNIPAM polymer chain and explore the influence of its elongation, stereochemistry, and temperature on the solute binding affinities. While we find that the excess adsorption generally raises with the size of the solute, the temperature-dependent affinity to the chains is highly solute specific and has a considerable dependence on the polymer elongation (\ie, polymer swelling state). We elucidate the molecular mechanisms of the selective binding in detail and eventually present how the results can be extrapolated to macroscopic partitioning of the solutes in swollen polymer architectures, such as hydrogels. 
\end{abstract}
\maketitle

	
\thispagestyle{plain}
\pagestyle{plain}

\pagenumbering{arabic}
\setlength\arraycolsep{2pt}

\section{Introduction}
\begin{figure*}[h!]\begin{center}
\begin{minipage}[b]{0.28\textwidth}\begin{center}
\includegraphics[width=\textwidth]{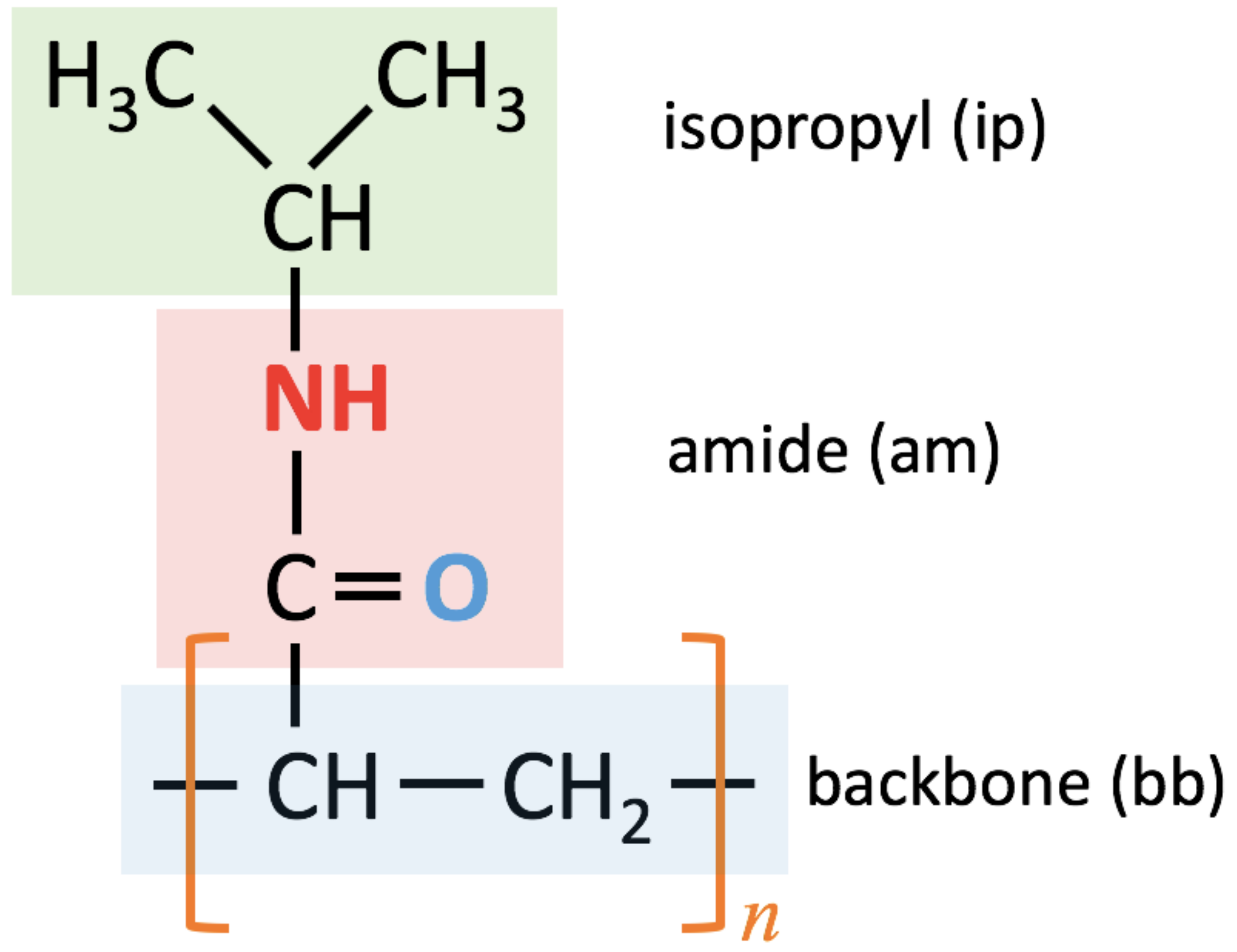} a)
\end{center}\end{minipage}\hspace{2ex}
\begin{minipage}[b]{0.3\textwidth}\begin{center}
\includegraphics[width=\textwidth]{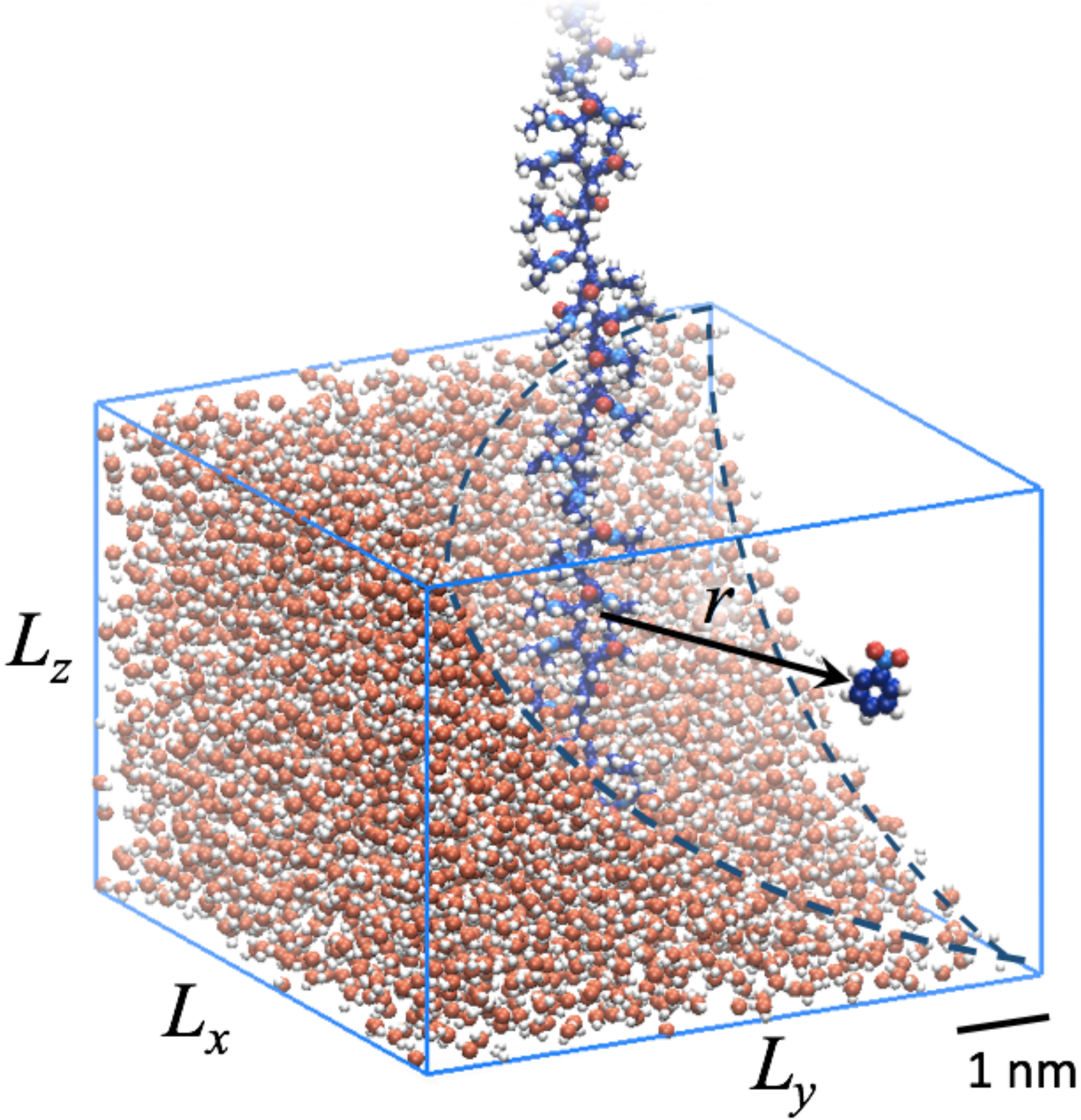} b)
\end{center}\end{minipage}\hspace{5ex}
\begin{minipage}[b]{0.34\textwidth}\begin{center}
\includegraphics[width=\textwidth]{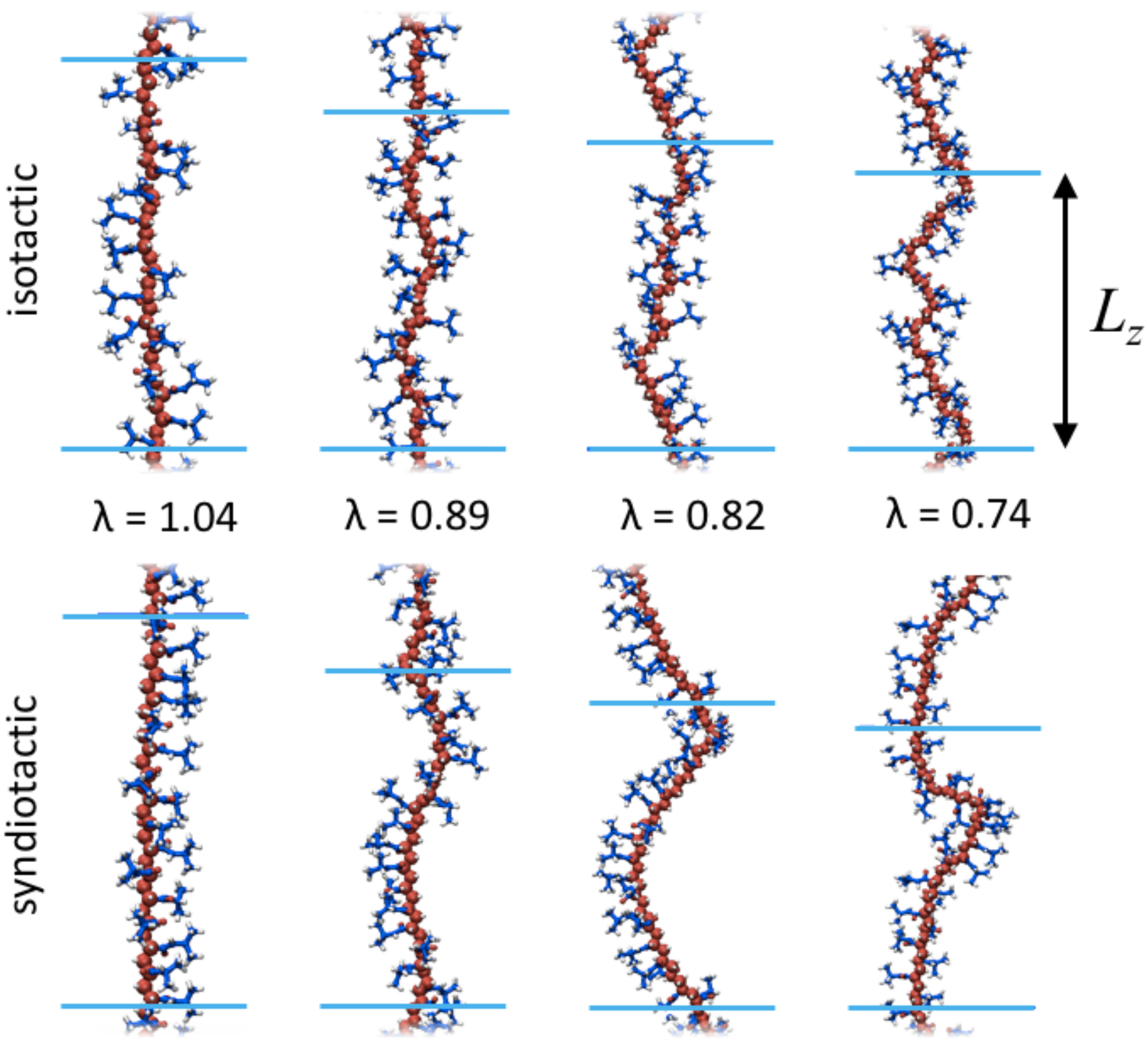} c)
\end{center}\end{minipage}
\caption{(a) Chemical structure of the PNIPAM polymer, featuring three functional groups. 
(b) Simulation snapshot: Periodically replicated isotactic PNIPAM chain composed of 20 monomers in the presence of a single nitrobenzene molecule. The lateral box sizes are $L_x$\,$=$\,$L_y$\,$=$\,6.6~nm, and the longitudinal box size is $L_z$\,$=$\,4.75~nm, corresponding to the relative extension of $\lambda$\,$=$\,0.89. Water is shown only partially in the simulation box (blue frame) for clarity.
(c) Snapshots of the periodic PNIPAM chain (water not shown) for different extensions $\lambda$ in the case of isotactic (top) and syndiotactic (bottom) stereochemistry. Horizontal lines denote the edges of the simulation box. The carbon atoms in the backbone are highlighted for better clarity.}
\label{fig:system}
\end{center}\end{figure*}

Thermoresponsive polymer architectures, such as brushes or microgels have gained substantial popularity in recent years due to their potential applications in the fields of drug delivery~\cite{stuartNatMat2010, kabanovAngew2009, oh2008development,yingApplicationsSM2011}, catalysis~\cite{jandtCatal2010, ballauffRoyal2009, dzubiellaAngew2012}, biosensing~\cite{yingApplicationsSM2011,ravaineGlucose2006}, thin-film techniques~\cite{yingApplicationsSM2011}, environmental science~\cite{serpeACS2011}, etc. 
Typical representatives in fundamental developments are polymer systems based on the popular poly(N-isopropylacrylamide) (PNIPAM) polymer (Fig.~\ref{fig:system}a), owing to its temperature-responsiveness close to room or body temperature and relatively simple chemistry~\cite{pelton2000, hudsonProgPolySci2004}. 
It has thus become a versatile active model component to push forward the development in soft material design, especially for responsive carrier systems~\cite{stuartNatMat2010, ballauffAngew2006, ballauffSmartPolymer2007}.

Inherent to all polymer-based carrier systems is that the adsorption and partitioning of molecules within the polymer matrix play a decisive role in its function, \eg, 
a well-controlled permeability window needs to be established for ligand uptake and release in drug delivery systems. Another important examples discussed with more focus in this work are active nanocarrier systems used in fluid-phase catalysis based on metal nanoparticles, termed also active `nanoreactors'~\cite{ballauffAngew2006, dzubiellaAngew2012, lis-marzan2008, stamm2014}.  Here, catalytic nanoparticles are embedded in a stimuli-responsive polymer gel. In a solution, partitioning and diffusion of various reactants through the gel can be controlled by external parameters such as temperature or pH. Using cross-linked PNIPAM hydrogel, which undergoes a volume phase transition at 32~$^{\circ}$C, it is possible to tune the permeability of the polymer matrix and the amount of reactants in the gel, and by that control the rate of the reaction itself~\cite{ballauffChemSocRev2012}.

Recently, selective catalysis by harvesting selective permeabilites of reactants has been realized in a simple model system of a single gold nanoparticle encapsulated in a hollow PNIPAM sphere, which enables a temperature-triggered switch of catalytic reactions~\cite{dzubiellaAngew2012}.  In that study, Wu et al.\ investigated the reduction of nitrobenzene~(NB) and 4-nitrophenol~(NP) by borohydride in aqueous solution into aminobenzene~(AB) and 4-aminophenol~(AP), respectively. They demonstrated that NB reacts much faster than NP at higher temperatures above the volume transition of PNIPAM, whereas the reduction of NP is preferred at lower temperatures, below the transition. These reactions have become benchmark model reactions to assess the catalytic activity of metal nanoparticles embedded in various carrier systems and nanoreactors~\cite{songAdvMat2008, dzubiellaAngew2012, ballauffChemSocRev2012}.
Hence, the key to control the catalytic properties of a responsive nanoreactor is to understand the partitioning and transport of the reactants through the polymer shell~\cite{dzubiellaAngew2012}. The problem is multifaceted and involves many intertwined elements: diffusion kinetics of reactants through the gel, their enrichment (absorption) in the gel, interaction with the polymer chains, the presence of cross-linkers and other cosolutes, reaction mechanisms at the catalytic surface, etc. Due to the high complexity of the problem, its understanding requires detailed studies of individual aspects before a universal picture can be established.
A key element to start with is first to understand the details of the solute--gel interactions on the atomic scale, attainable through atomistic computer modeling.
In recent years, a number of simulation studies focused on solute--polymer interactions that affect the conformation~\cite{smiatekPCCP2016}, solubility~\cite{heyda2013rationalizing} as well as critical solution temperature of a PNIPAM polymer~\cite{heydaTDHofmeister}. A particular attention was given to understanding of cononsolvency effects~\cite{kremerNatComm2014, vdVegtJPCB2015, vdVegtPCCP2015}, which appear with increasing solute concentrations~\cite{venzmerMacromolecules1990, muthuMacromolecules1991}.

In this work, we investigate adsorption characteristics of different solutes on a thermoresponsive PNIPAM polymer within a framework of all-atom explicit-solvent Molecular Dynamics (MD) simulations. We model an extended PNIPAM polymer chain and study the adsorption of representative classes of chemical compounds 
found in small ligands and reactants with an emphasis on NB and NP, motivated by the model catalytic experiments~\cite{dzubiellaAngew2012}. 
We examine the conformational, stereochemical, 
temperature, as well as concentration
influences on the solutes binding affinities and elucidate the details of 
the molecular mechanisms of binding. At the end, we quantitatively relate the adsorption coefficients of a single chain obtained from the simulations to the solute partitioning and the solvation free energy of a swollen hydrogel. 

\section{Methods}
\subsection{Atomistic polymer model}

In the present study, we investigate the adsorption properties of various solutes on an elongated PNIPAM polymer chain. 
A single isolated and elongated PNIPAM chain can be considered as a polymer sequence in a diluted brush or hydrogel in the swollen state, where the adjacent chains are far apart and do not interact with each other.
To that end, we set up an all-atom model of a single elongated PNIPAM chain in explicit water.
We enforce cylindrical symmetry of the 20-monomer-long chain by stretching and periodically replicating it along the $z$-direction of the rectangular simulation box~\cite{horinekJPCA2011, heyda2012ion-specific}, see Fig.~\ref{fig:system}b for a simulation snapshot. This well-defined set-up enables us to analyze and evaluate water and solute distributions around the chain as a function of radial distance $r$. 
{By imposing the elongated configuration we suppress larger conformational changes and bends of the chain, which allows us to explore the principal polymer--solute interactions relevant for the discussion on swollen hydrogels.}

For the PNIPAM as well as the solute molecules we use the all-atom OPLS-AA~\cite{opls1988} force field and the SPC/E water model~\cite{spce}. This combination of force fields reproduces the experimentally measured thermoresponsive properties of PNIPAM considerably well~\cite{walter-PNIPAM2010, vegtJPCB2011}.
The simulations are performed using the GROMACS~5.1 simulation package~\cite{gromacs,gromacs2013}. We choose the canonical constant-pressure (NPT) ensemble where only the lateral box sizes ($L_x$\,$=$\,$L_y$\,$\approx$\,6.5~nm) are symmetrically adjusted in order to maintain the external pressure of 1~bar via Berendsen barostat~\cite{berendsenT}. The height $L_z$  of the box, on the other hand, is kept fixed, which sets the longitudinal extension of the chain, \Fig~\ref{fig:system}c.
The system temperature is maintained by the velocity-rescaling thermostat~\cite{v-rescale} with a time constant of 0.1\,ps.
The Lennard-Jones (LJ) interactions are cut off at $r_\textrm c=1.0$~nm. Electrostatics is treated using Particle-Mesh-Ewald (PME) methods~\cite{PME1,PME2} with a 1.0~nm real-space cutoff.
Prior to the production runs, the systems are equilibrated for at least 1~ns. Production runs for sampling solutes adsorption have a duration of 1000--2000~ns.

\subsection{Solutes}
\begin{figure*}[t!]\begin{center}
\begin{minipage}[b]{0.85\textwidth}\begin{center}
\includegraphics[width=\textwidth]{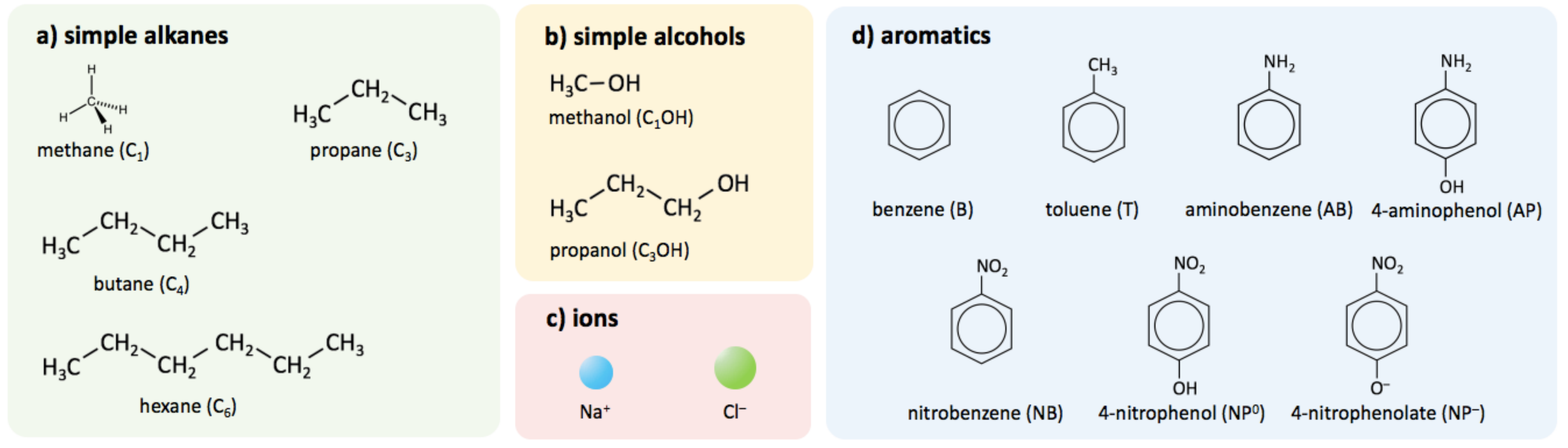}
\end{center}\end{minipage} 
\caption{The solutes studied in our simulations: (a)~Simple alkanes, (b)~simple alcohols, (c)~ions, and (d)~aromatic molecules.}
\label{fig:molecules}
\end{center}\end{figure*}

{We are primarily interested in adsorption properties in the infinite dilution limit of a solute, that is, when the solute bulk concentration is low enough that the collective effects of the solute molecules play negligible role. 
We will examine the effects of elevated concentrations for the cases of nitrobenzene and nitrophenolate. In the case of infinite dilution limit, we simulate the PNIPAM chain in water in the presence of a single solute molecule, as shown in  \Fig~\ref{fig:system}b.}
We examine four different categories of solutes: simple alkanes, simple alcohols, monovalent ions, and aromatic molecules (all shown in \Fig~\ref{fig:molecules}).

Motivated by the nanocatalytic benchmark experiments~\cite{dzubiellaAngew2012}, we devote special attention to two reactants, nitrobenzene~(NB) and 4-nitrophenol~(NP). In an aqueous environment, the acidity constant of the hydroxyl (OH) group in NP is p$K_\trm a$\,$=$\,7.15~\cite{nitrophenol-pKa}. Therefore, already under neutral conditions the hydroxyl group partially deprotonates and forms a nitrophenolate (NP$^-$) anion (\Fig~\ref{fig:molecules}d). Under alkaline conditions governing the catalytic experiments~\cite{dzubiellaAngew2012}, with pH value approaching 10, NP becomes entirely deprotonated.
In our simulations we therefore study both variants, the neutral, protonated~(NP$^0$) molecule as well as the charged, deprotonated~(NP$^-$) ion.
The products of the reductions of NB and NP are respectively aminobenzene~(AB) and aminophenol~(AP), which we also include in our study.

For the solute molecules we use the OPLS-AA force field~\cite{opls1988, priceOPLS2001}, the same as for PNIPAM.
Aromatic molecules are conjugated systems, with $\pi$-electrons causing long-range electronic interactions across the aromatic ring. This means that the electron distribution in an entire aromatic molecule can be influenced by attached residues, such as nitro and hydroxyl groups. 
Since the OPLS force field does not explicitly consider mesomeric effects in its partial-charge parameters, we verify the results obtained by the OPLS force field by an additional set of  simulations where we implement the partial charges based on explicit quantum-mechanical~(QM) calculations. We perform the QM calculations by use of the Gaussian~09 software package~\cite{g09} as explained in more detail in the \SItext.
In the QM-based parameterization, we adopt all other model parameters of the solutes from the original OPLS-AA force field. In the following, we refer to `OPLS' and `QM' parameterizations for the solute partial charges based on the OPLS-AA force field and our QM calculations, respectively. 

The partial charges of NP$^-$, on the other hand, are not comprised within the OPLS force field and need to be resolved subsequently. In our study, we employ two different approaches for determining the partial charges of NP$^-$.
In the first approach, we determine the partial charges based on QM calculations using the Gaussian software. In the second approach, we start with the charge distribution of the OPLS-based partial charges of NP$^0$ on top of which we add a difference between the charge distributions in NP$^-$ and NP$^0$ evaluated by the QM calculations using Gaussian. The latter hybrid approach we term as `OPLS/QM' parameterization. For details see the \SItext.
Note that our QM method differs from the method used in the OPLS force field, where the partial charges for nitro molecules are based on an initial ab-initio guess and subsequently fine-tuned to yield acceptable conformational energetics and pure liquid properties~\cite{priceOPLS2001}. The rationale behind the OPLS/QM parameterization is that the fine tuning more likely remains retained, since the difference in electronic densities between two structures is more likely independent of a method used.


\subsection{Tacticity and polymer elongation}
The carbon atoms in the PNIPAM backbone bearing side chains are chiral, giving rise to two possible enantiomers for each monomeric unit in the chain. The sequence of the two kinds of monomers determines the tacticity of a polymer.
In our simulations, we study two cases of tacticity, the isotactic (meso-diad) PNIPAM polymer, with all the side chains located on the same side of the backbone, and the syndiotactic (racemo-diad) one, where the side chains alternate in the position along the chain.
Experimentally it is known that tacticity, which can be controlled by polymerization procedures, significantly influences the hydration properties and volume transition temperature of the PNIPAM polymer~\cite{ray_Polymer2005, katsumotoMacromol2008}. The influence of the tacticity on the transition temperature has been reported also in computer simulations~\cite{stevens_Macro2012}.
As a matter of fact, the isotactic and syndiotactic stereochemistries represent two extreme scenarios in hydration behavior, such as transition temperature and swelling ratio. Other combinations, such as atactic polymer, where both variants of stereoisomeric units are randomly distributed along the chain, exhibit intermediate properties~\cite{ray_Polymer2005, takashiJPS2006, biswasLangmuir2010}.

The conformational property in our model is the longitudinal extension of the polymer in the $z$ direction.
{
We define an elongation $\lambda$ as the ratio between the extension of the chain in the $z$ direction and its corresponding contour length $L_\trm c$. In our simulations, it is equal to the ratio of the box height $L_z$ to the contour length $L_\trm c$ of the 20-monomer-long sequence of the chain}, 
\begin{equation}
\lambda=\frac{L_z}{L_\trm c}.
\label{eq:lambda}
\end{equation}
For $\lambda$$>$1, the chain is overstretched, while for $\lambda$$<$1 it becomes loose with larger fluctuations in the lateral ($x, y$)-directions.
From independent simulations of a finite, non-replicated 20-monomer-long PNIPAM chain, we determine the contour length per monomer--monomer distance to be $\Delta L_\textrm c$\,$=$\,0.266 and 0.264~nm for the isotactic and syndiotactic chains, respectively (see the \SItext).  
Figure~\ref{fig:system}c shows simulation snapshots of the replicated isotactic and syndiotactic PNIPAM  chains for different extension parameters $\lambda$. The blue horizontal lines denote the edges of the simulation box. The backbone carbon atoms are displayed by red spheres for better visibility. As seen, a significant difference between the isotactic (top) and syndiotactic (bottom) variants appears in the chain conformations. The isotactic chain shows locally a more curly nature of the backbone compared with the locally smoother syndiotactic one. 


Due to the cylindrical {symmetry} of the  system, it is convenient to use the two-dimensional radial distribution function~(RDF) of a solute and water molecules around the polymer to analyze the adsorption properties~\cite{horinekJPCA2011, heyda2012ion-specific}. 

\section{Results and discussion}
\begin{figure}[t]\begin{center}
\begin{minipage}[b]{0.45\textwidth}\begin{center}
\includegraphics[width=\textwidth]{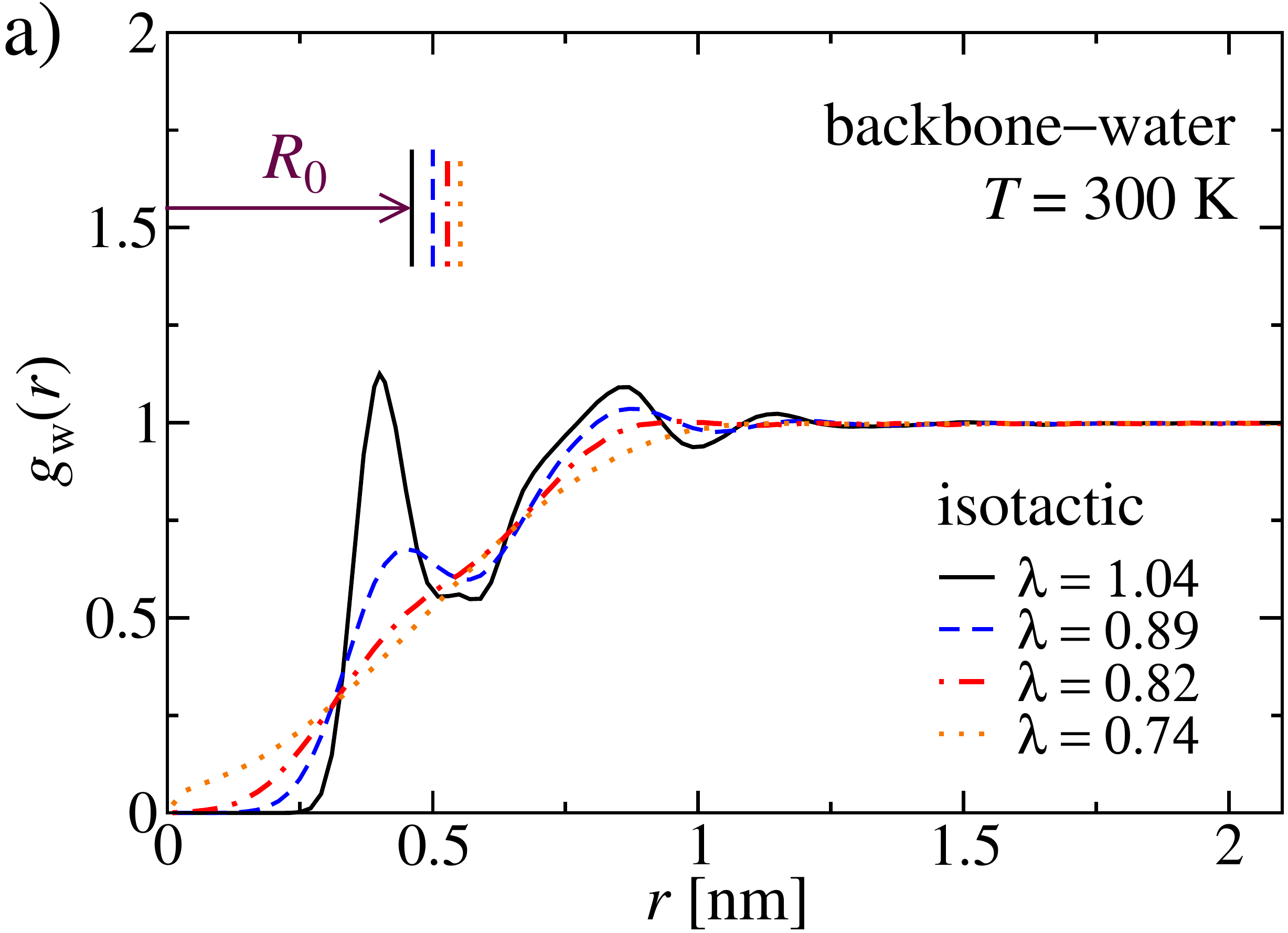}
\end{center}\end{minipage} 
\begin{minipage}[b]{0.45\textwidth}\begin{center}
\includegraphics[width=\textwidth]{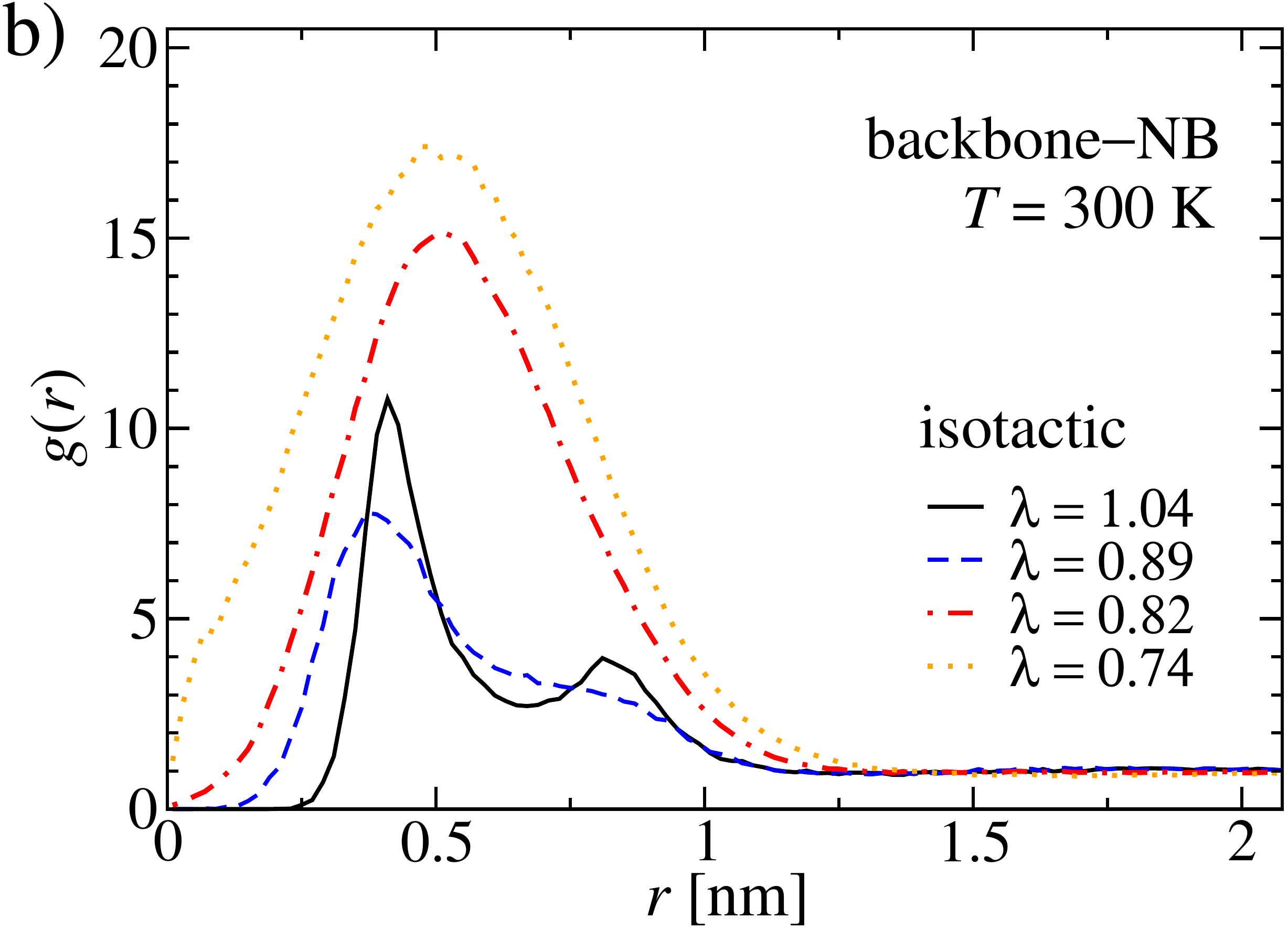}
\end{center}\end{minipage} 
\caption{(a) Cylindrical radial distribution function of backbone--water for four different extension parameters $\lambda$ of the isotactic PNIPAM at $T$\,$=$\,300~K. The radii of the Gibbs dividing surface $R_0$  are shown by vertical bars for different values of $\lambda$. (b)~Cylindrical RDF of backbone--NB for the same systems as shown in~(a). The RDF is computed only for the carbon atoms in the aromatic ring.}
\label{fig:RDF-NB}
\end{center}\end{figure}

In the first step, we analyze the water structure around the chain. 
The cylindrical RDF of backbone--water, $g_\trm w(r)$, plotted in \Fig~\ref{fig:RDF-NB}a, shows the radial water distribution for four different extensions $\lambda$ around the isotactic PNIPAM chain. The largest extension of $\lambda$\,$=$\,1.04 exhibits profound peaks in the RDF, indicating layering of water molecules around the elongated polymer chain. Upon loosening the chain (\ie, decreasing $\lambda$), increased backbone fluctuations suppress the layering and the peaks disappear.
We determine the Gibbs dividing radius $R_0$ in cylindrical coordinates as~\cite{horinekJPCA2011}
\begin{equation}
R_0^2=2\int_0^\infty[1-g_\textrm w(r)]\,r\rmd r,
\label{eq:R0}
\end{equation}
and represents an effective radius of the polymer chain.
Its value ranges from $R_0$\,$=$\,0.46~nm for a very elongated chain ($\lambda$\,$=$\,1.04) to $R_0$\,$=$\,0.55~nm for a loose chain ($\lambda$\,$=$\,0.74), see \Fig~\ref{fig:RDF-NB}a. 

Similarly as for backbone--water molecules,  we evaluate the cylindrical RDF for backbone--solute molecules, $g(r)$.
An example in \Fig~\ref{fig:RDF-NB}b shows the RDFs of the aromatic ring in a NB molecule for the same conditions as in (a).
A very elongated chain (with $\lambda$\,$=$\,1.04) gives rise to a structured RDF with two clearly distinct peaks at 0.4~nm and~0.8 nm, respectively. The first peak at smaller $r$ corresponds to an adsorption of the molecule on the backbone, whereas the second one corresponds to the adsorption on the side chains. For a less extended chain, the pronounced fluctuations of the backbone smear out the profile and the two peaks in the RDF fuse together into a single peak. 

The RDFs enable the evaluation of the amount of adsorbed solute on the polymer. The number of adsorbed solute molecules $\Gamma$ on the polymer chain with a longitudinal length $L_z$ can be obtained in a standard way by integrating the excess of the solute density $c(r)$\,$=$\,$c_0 g(r)$, with $c_0$ being the bulk solute concentration, along the spatial coordinates~\cite{horinekJPCA2011},
\begin{equation}
\Gamma=L_z \int_0^{R_0} c(r) 2\pi r\rmd r +L_z \int_{R_0}^\infty [c(r)-c_0] 2\pi r\rmd r.
\end{equation}
The adsorbed solute amount $\Gamma$ is proportional to the longitudinal chain length $L_z$ as well as, in the infinite dilution limit, to the bulk solute concentration $c_0$. The adsorption can be expressed in a compact form as
\begin{equation}
\Gamma=\Gamma' c_0 L_z,
\label{eq:GammaTOT}
\end{equation}
where we introduced the {adsorption coefficient} $\Gamma'$
\begin{equation}
\Gamma'=2\pi\int_0^\infty[g(r)-1]r\rmd r+\pi R_0^2,
\label{eq:Gamma1}
\end{equation}
which does not explicitly depend on $L_z$ and $c_0$. {Note that at higher concentrations, solute--solute interactions can become important and therefore $\Gamma'$ can depend on $c_0$, as we will demonstrate later.}
The second term stems from the integration over the Gibbs dividing surface. Given the range of $R_0$ of 0.46--0.55~nm, its contribution $\pi R_0^2\!\approx$\,0.7--1.0~nm$^2$ is typically much smaller than the overall {adsorption coefficient} $\Gamma'$.
{
Error estimates for the adsorptions are obtained by the block averaging method, where the simulation trajectory is divided into 5--10 blocks and $\Gamma'$ is calculated for each block.
This also enables us to demonstrate that the single solute particle adsorbs and desorbs from the chain several times during the simulation, which confirms that ergodicity is operative and that simulation lengths (1000--2000~ns) are sufficient.}

\begin{figure}\begin{center}
\begin{minipage}{0.45\textwidth}\begin{center}
\includegraphics[width=\textwidth]{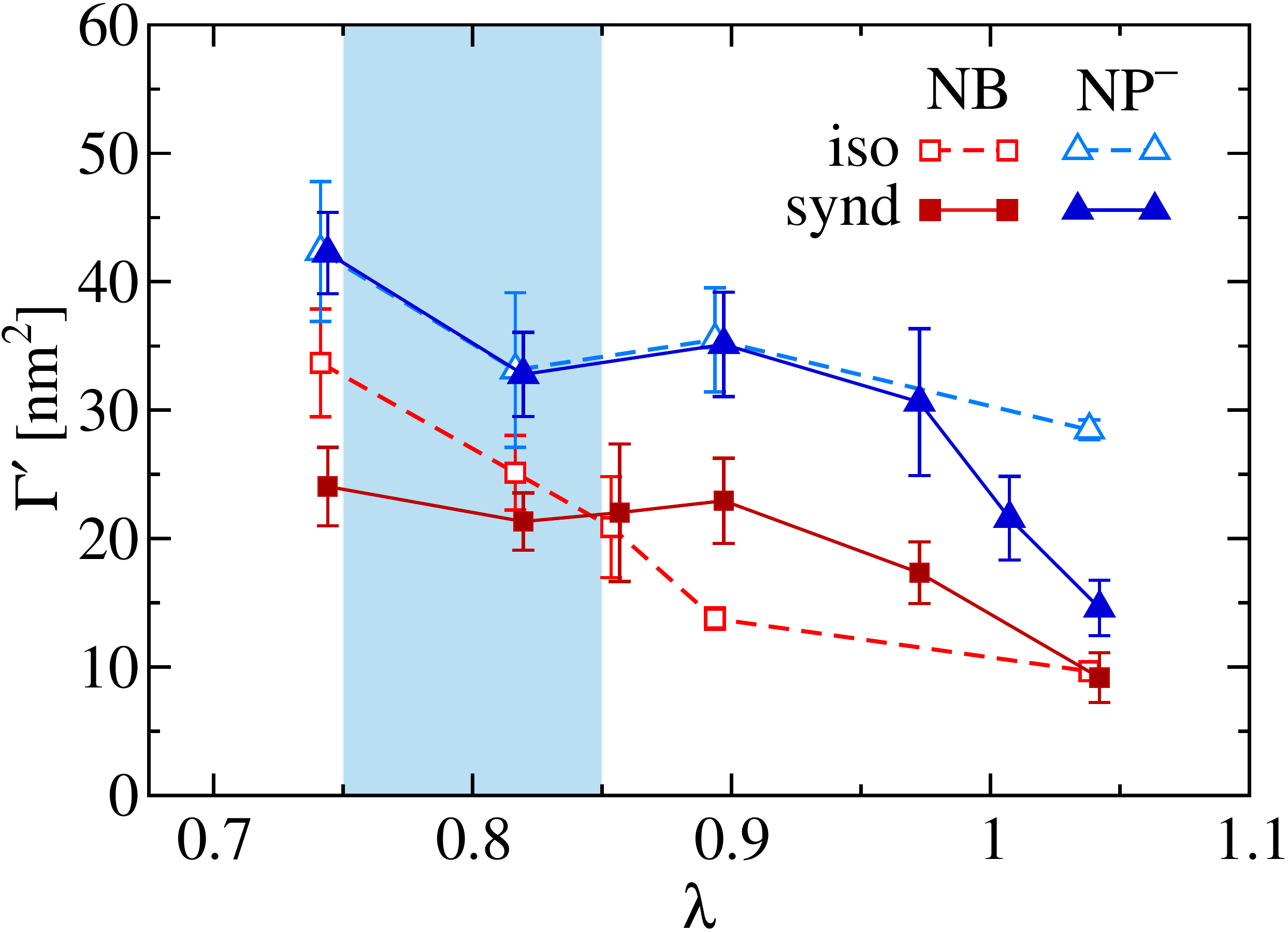}
\end{center}\end{minipage}
\caption{{Adsorption coefficients} of NB and NP$^-$ on a PNIPAM chain as a function of its extension parameter $\lambda$\,$=$\,$L_z/L_\trm c$ at $T$\,$=$\,300~K. Void symbols correspond to the isotactic and full symbols to the syndiotactic PNIPAM chain. The blue shaded region $\lambda$\,$=$\,0.75--0.85 serves as an orientation for the expected maximal chain extension in a swollen PNIPAM hydrogel.}
\label{fig:Gamma-Lz}
\end{center}\end{figure}

Figure \ref{fig:Gamma-Lz} shows the {adsorption coefficients} $\Gamma'$ of NB and NP$^-$ on the isotactic and syndiotactic PNIPAM chains as a function of their extension parameters $\lambda$. 
The  adsorption significantly depends on the extension of the chain as well as its tacticity.
Quite generally, the adsorption on a very extended chain is lower than on a loose one.
The dependence partially comes from the definition of $\Gamma'$, because it is defined per projected longitudinal length. That means that the smaller the extension, the more polymer (\ie, more monomeric units) is contained per given longitudinal length.
In the \SItext\ we show a similar plot expressed per chain's contour length $L_\trm c$, that is $\lambda \Gamma'$. Even in that case, the adsorption tends to decrease, albeit less, with the extension, especially for 
a very stretched and even overstretched chain. 
The reason for this behavior most probably lies in an intricate interplay between the orientation of the side chains and the solute. As can be seen, in all cases the adsorption of NP$^-$ is  {slightly} higher than for NB. We will discuss the possible reasons for this difference later on.

In a typical swollen PNIPAM hydrogel, polymer chains spanning between adjacent
cross-linkers are predominantly extended. The degree of extension depends on various parameters, such as the degree of cross-linking and temperature. We anticipate that cross-linked chains in a swollen gel that is in a mechanical equilibrium experience neither compressive nor tensile  stress on average~\cite{caiMechanics2011}.
In order to obtain an estimation for the elongation of chains in a typical dilute and swollen PNIPAM hydrogel, we analyze the end-to-end distance of a single non-replicated PNIPAM chain in water, see the \SItext. At low temperatures, relevant for the swollen state, a single chain exhibits maximal extensions of around $\lambda$\,$=$\,0.75--0.85, which is indicated by a blue shaded region in \Fig~\ref{fig:Gamma-Lz}.
We will base our further analyses on the isotactic chain with the representative extension of $\lambda\!=\,$0.82.

\subsection{Adsorption of different solutes}
We now compare the adsorptions of four different classes of chemical compounds on the PNIPAM chain: aromatic molecules, simple alkanes, simple alcohols, and ions, all listed in \Fig~\ref{fig:molecules}.
The comparison allows us to gain important insights into general mechanisms of adsorption, such as how do molecular size and different functional groups contribute to the adsorption propensity.

\begin{figure*}[t]\begin{center}
\begin{minipage}[b]{0.93\textwidth}\begin{center}
\includegraphics[width=\textwidth]{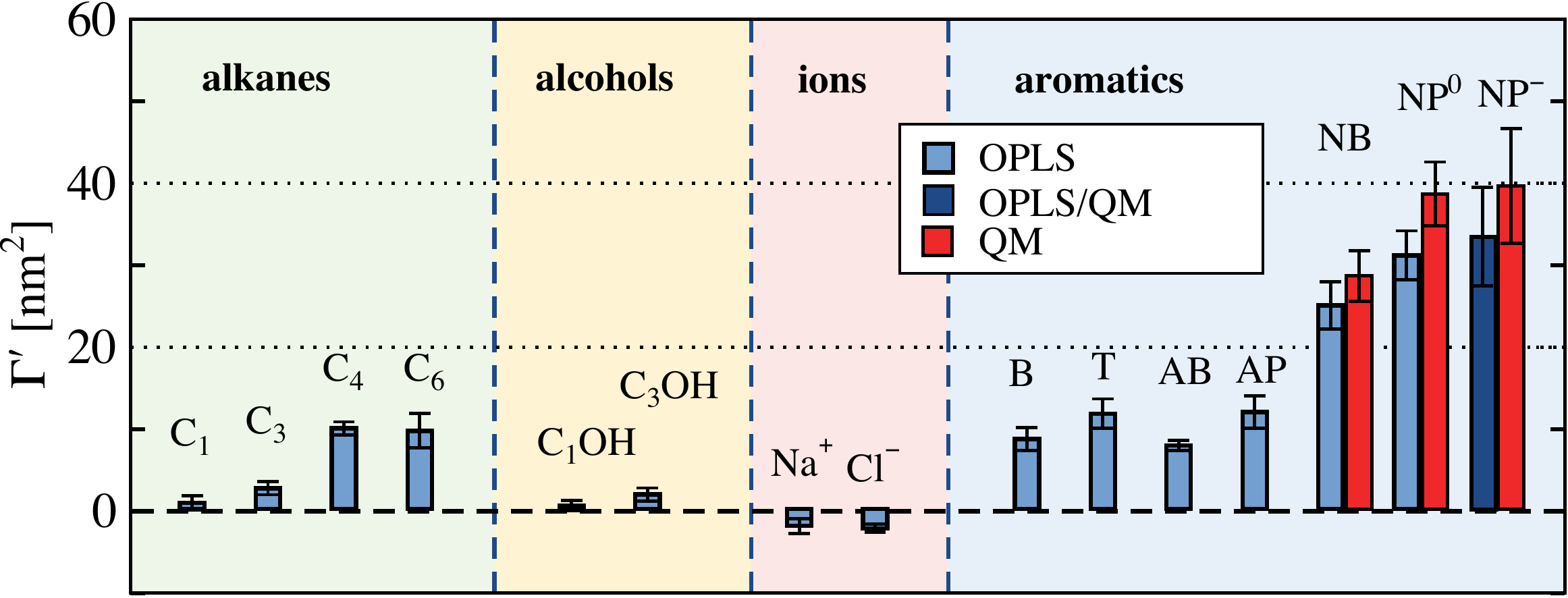}
\end{center}\end{minipage}
\caption{{Adsorption coefficients} of various types of reactants on the isotactic PNIPAM chain with the extension $\lambda$\,$=$\,0.82 at $T$\,$=$\,300~K. See \Fig~\ref{fig:molecules} for a description of the molecules and the \SItext\ for the details of the force fields.}
\label{fig:all}
\end{center}\end{figure*}

Figure \ref{fig:all} shows the adsorption of all the simulated solutes on the isotactic PNIPAM chain with an extension of $\lambda\!=$\,0.82 at $T\!=$\,300~K.
Starting with linear alkanes, we note that their relatively low {adsorption coefficients} $\Gamma'$ rise with their molecular size. 
For larger molecules,  more water can be liberated at adsorption at the hydrophobic PNIPAM sites, leading to larger overall binding affinities.  
The hydroxylation, that is, adding a polar hydroxyl~(OH) group and by that transforming purely hydrophobic alkanes into polar alcohols, does not change adsorptions significantly.  The two ion types in our case,  sodium cation Na$^+$ and chloride anion Cl$^-$, are both repelled from the polymer. Quite generally, most of the simple ions (often classified as `kosmotropes') have the tendency of being strongly hydrated in water  and are therefore repelled from a low-dielectric environment, such as the PNIPAM polymer~\cite{cremerJACS2005, heydaTDHofmeister}.

An illuminating trend is observed for aromatic molecules. Benzene (B), the simplest among aromatic molecules, exhibits relatively low adsorption, comparable to similarly-sized alkanes, such as butane (C$_4$) or hexane (C$_6$).
Toluene (T), which additionally has one hydrophobic methyl group attached to the benzene ring, has slightly larger {adsorption coefficient}, consistent with the observed size-dependence trend in alkanes. 
Aminobenzene~(AB), which is a benzene with an additional amino~(NH$_2$) group, displays essentially the same adsorption as benzene.
The adsorption is not significantly affected upon hydroxylation, by that transforming it into  4-aminophenol~(AP).
A dramatic change occurs when a nitro (NO$_2$) group is introduced to the benzene ring.
The adsorption of NB is around three times larger than that of B and AB. This means that a nitro group significantly facilitates the adsorption of the molecule with its high binding affinity to PNIPAM.
Finally, hydroxylating the NB molecule, which results in NP$^0$, does not change the overall adsorption considerably.
This observation is consistent with the observation in the comparison of alkanes vs.\ alcohols: 
Hydroxylating a molecule, and by that making it more polar, has an insignificant effect on the adsorption on a PNIPAM polymer.
As can be seen, nitrophenol has very similar adsorption affinities in the protonated (NP$^0$) and deprotonated (NP$^-$) states. Despite being an ion, the excess charge in NP$^-$ is not localized on the deprotonated oxygen O$^-$ atom but distributed over the entire molecule instead (see the \SItext). Consequently, the NP$^-$ ion behaves very different than small atomic ions like Na$^+$ and Cl$^-$.
This follows the well-accepted trend where bulkier monovalent ions, such as iodide, tend to be less hydrated and therefore do not show the typical dielectric repulsion from a low-dielectric medium~\cite{konovalovPRL2007, horinekCPL2009}.

In light of the nanoreactor-based catalysis the results suggest that the reduction products (AB and AP) accumulate in a PNIPAM hydrogel to much lesser extent than the reactants (NB and NP) and hence leaving the nanoreactor faster.

The results for the nitro molecules in \Fig~\ref{fig:all} stemming from the two different parameterizations of partial charges (OPLS and OPLS/QM vs.\ QM)  differ only by the extent of the error bars. This validates the use of both paramererizations for our purposes.

\subsection{Binding sites}
Further important insights into the adsorption of NB and NP can be obtained by determining the  sites  that are responsible for binding.
The integration of the RDF provides us only with the overall amount of adsorption $\Gamma$ of a solute on the chain. 
In the following, we analyze which chemical groups in a solute and PNIPAM tend to bind together most preferably.
\begin{figure*}[t]\begin{center}
\begin{minipage}[b]{0.95\textwidth}\begin{center}
\includegraphics[width=\textwidth]{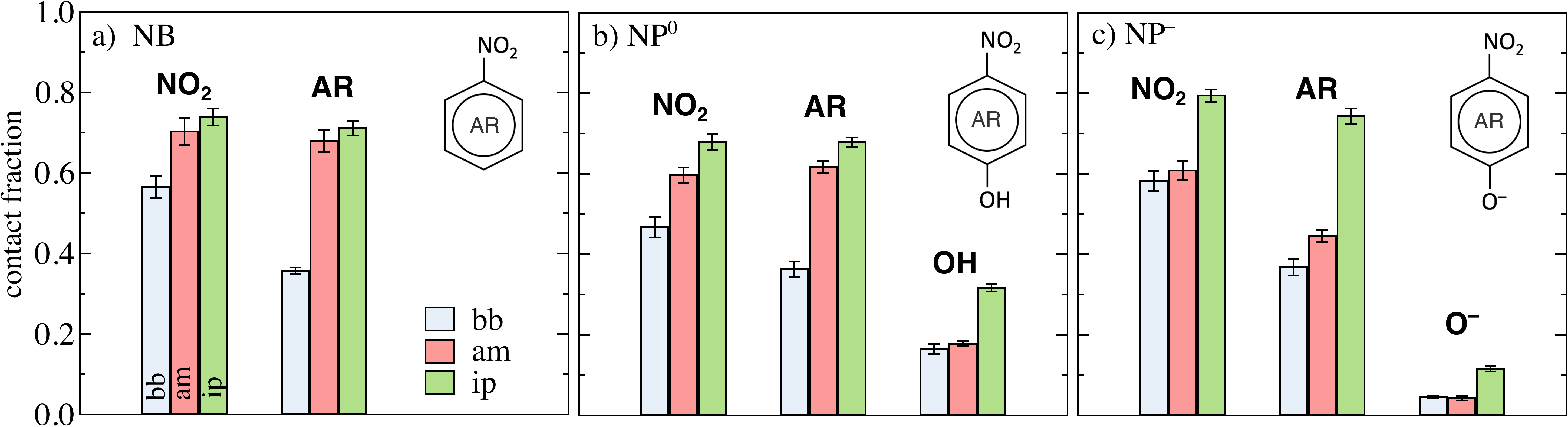}
\end{center}\end{minipage} 	
\caption{Contact fractions of different groups in a bound solute: (a) NB, (b) NP$^0$, and (c) NP$^-$
  to the specific locations on an isotactic PNIPAM chain with an extension $\lambda$\,$=$\,0.82 and at $T\!=$\,300~K: backbone (bb), amide (am), and isopropyl (ip). 
  See also \Fig~\ref{fig:system}a for a schematic illustration of the groups in PNIPAM.
  The force fields used were OPLS for NB and NP$^0$, and OPLS/QM parameterization for NP$^-$ (see the \SItext).
}
\label{fig:fractions}
\end{center}\end{figure*}

In the PNIPAM polymer, we distinguish three different functional groups: the backbone~(bb), amide groups~(am), and isopropyl groups~(ip), as schematically depicted in \Fig~\ref{fig:system}a.
In the NB molecule we distinguish between the aromatic ring~(AR) and the nitro group~(NO$_2$), whereas NP additionally possesses a hydroxyl group, either protonated~(OH) or deprotonated~(O$^-$), \Fig~\ref{fig:molecules}d.

We define a contact fraction of two specific groups (one in the solute and the other in the polymer) as the ratio between the time the two groups are in contact and the overall time the solute molecule is bound to the PNIPAM chain.
The details are described in the \SItext.
\Figure~\ref{fig:fractions} shows the contact fractions of different functional groups in NB, NP$^0$, and NP$^-$ to specific components of the isotactic PNIPAM chain with an extension $\lambda$\,$=$\,0.82 at $T$\,$=$\,300~K.
All three molecules bind most preferably to the isopropyl~(ip) groups. The main reason lies in the fact that the isopropyl groups are also the most accessible parts of the chain. The contact fractions with the other two, less accessible groups (bb and am), are accordingly smaller.
As can be seen for all three molecules, the nitro group shows a somehow larger tendency of adsorption on the backbone than the aromatic ring and the hydroxyl group.

The most striking observation is that the hydroxyl group, be it protonated~(OH) or deprotonated~(O$^-$), has significantly smaller affinity to the PNIPAM than the other parts of the nitrophenol molecule. This trend is also consistent with the observation in \Fig~\ref{fig:all}, that the additional hydroxyl group minimally contributes to the overall adsorption.
Particularly the deprotonated hydroxyl group O$^-$ in NP$^-$ experiences extremely low contact fraction and barely forms any association with PNIPAM.
The low binding propensity cannot be explained solely by the electrostatic repulsion from the low dielectric medium of the polymer, because, as already mentioned, the overall excess charge in the NP$^-$ ion is distributed throughout the whole molecule.
The low contact fraction of the deprotonated hydroxyl group thus suggests that it is well hydrated and it does not give away its hydration shell readily.
A nitrophenol molecule therefore preferentially binds to the PNIPAM with the NO$_2$ group oriented towards the chain and sticking the hydroxyl group away from it.
Even though the hydroxyl terminus does by itself directly not contribute significantly to the binding, it affects it through the nitro group on the other end of the molecule. 
Since the nitro group has a negative mesomeric ($-$M) effect, it takes up significant electron density via $\pi$-electron conjugation effects from the hydroxyl group with a positive mesomeric ($+$M) effect~\cite{exnerNitroChemSocRev1996}, which effects its polarity (compare the partial charge distributions in the \SItext). 

The binding analysis brings us to an important conclusion: a more polar 4-nitrophenol reactant 
 exhibits only {slightly} larger adsorption to a PNIPAM chain compared with a less polar nitrobenzene. The binding is predominantly governed by the hydrophobic effect of the aromatic ring (AR) and the polar nitro group (NO$_2$).

\subsection{Temperature dependence}
Temperature is one of the key control parameters in responsive carrier materials, predominantly due to conformational thermoresponsive properties of the polymer architecture.
By performing the simulations at different temperatures, we evaluate the temperature dependence of solute adsorptions on an elongated PNIPAM chain. 
\begin{figure}[h]\begin{center}
\begin{minipage}[b]{0.48\textwidth}\begin{center}
\includegraphics[width=\textwidth]{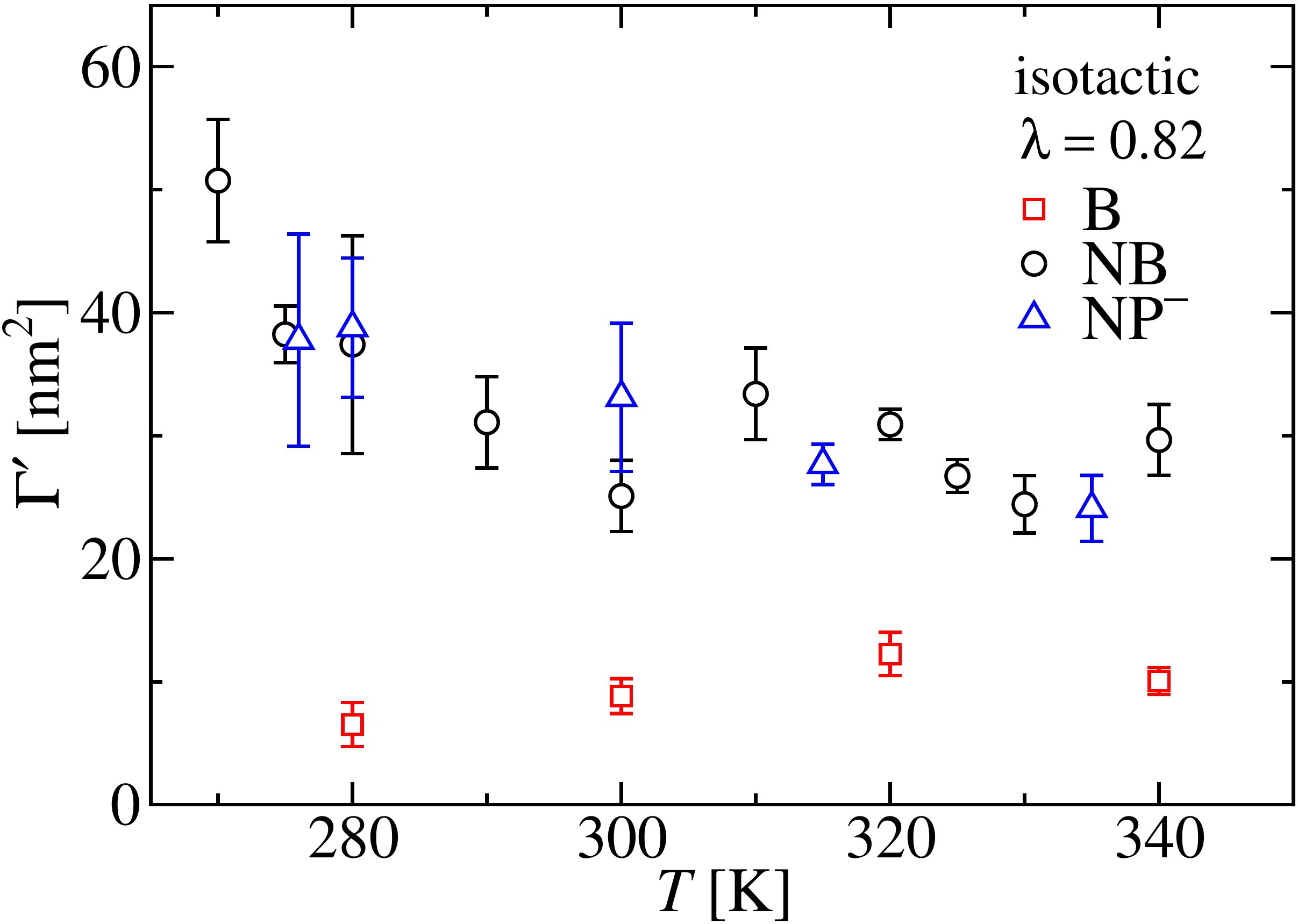}
\end{center}\end{minipage} 
\caption{Temperature-dependent {adsorption coefficients} of aromatic solutes on an isotactic elongated PNIPAM chain with an extension $\lambda$\,$=$\,0.82. 
{Note that above the transition temperature ($T$\,$>$\,305~K) the given elongation can be achieved only for a constrained chain, since an unconstrained chain undergoes the coil-to-globule collapse.}}
\label{fig:GammaT}
\end{center}\end{figure}

Figure \ref{fig:GammaT} shows {adsorption coefficients} of B, NB, and NP$^-$.
The relevant temperature range of an {unconstraint} PNIPAM hydrogel in the swollen state is $T$\,$<$\,305~K. The adsorption of benzene, which is essentially driven by the hydrophobic effect, slightly increases with increasing temperature. 
This can be attributed to the entropic origin of the hydrophobic effect.
On the other hand, both nitrated aromatics, NB and NP$^-$, exhibit the opposite trend: the adsorption {coefficient} drops with temperature, 
indicating a dominating enthalpic effect. However, we see that for a given enforced polymer elongation {(which prevents the chain to collapse above the transition temperature)}, the temperature has only moderate effect on the adsorption properties. For instance, in the temperature interval 280--340~K the adsorption {coefficient} gradually drops by around 30--40\% for both nitro compounds.
However, in a PNIPAM hydrogel, which undergoes the volume transition, the adsorptions are expected to change abruptly and dramatically when passing the transition temperature and all polymers collapse and generate an `oily' matrix quite different in its properties than isolated polymer chains. A substantial difference of partitioning in a PNIPAM gel below and above the transition has been demonstrated for the case of methyl orange~\cite{molinaPolymer2012,jiaJMCA2016}.

\subsection{Dependence of adsorption on solute concentration}
{
In this study, we are primarily focused on the infinite dilution limit of the solute, as is typically the case for reactants in nanocatalysis~\cite{dzubiellaAngew2012,ballauffChemSocRev2012,jiaJMCA2016}.
  In this regime, the solute--solute interactions are negligible and the solute adsorption on a polymer $\Gamma$ is proportional to the bulk solute concentration $c_0$.
Generally, the linear trend is expected to break down at larger concentrations, which we will briefly address in this section for the cases of NB and NP$^-$. To this end, we perform simulations of the PNIPAM chain in the presence of up to eight solute particles in the simulation box. For the case of NP$^-$ we add neutralizing Na$^+$ counterions, which approximately corresponds to the scenario of low salt concentration. }

\begin{figure}[h!]\begin{center}
\begin{minipage}{0.45\textwidth}\begin{center}
\includegraphics[width=\textwidth]{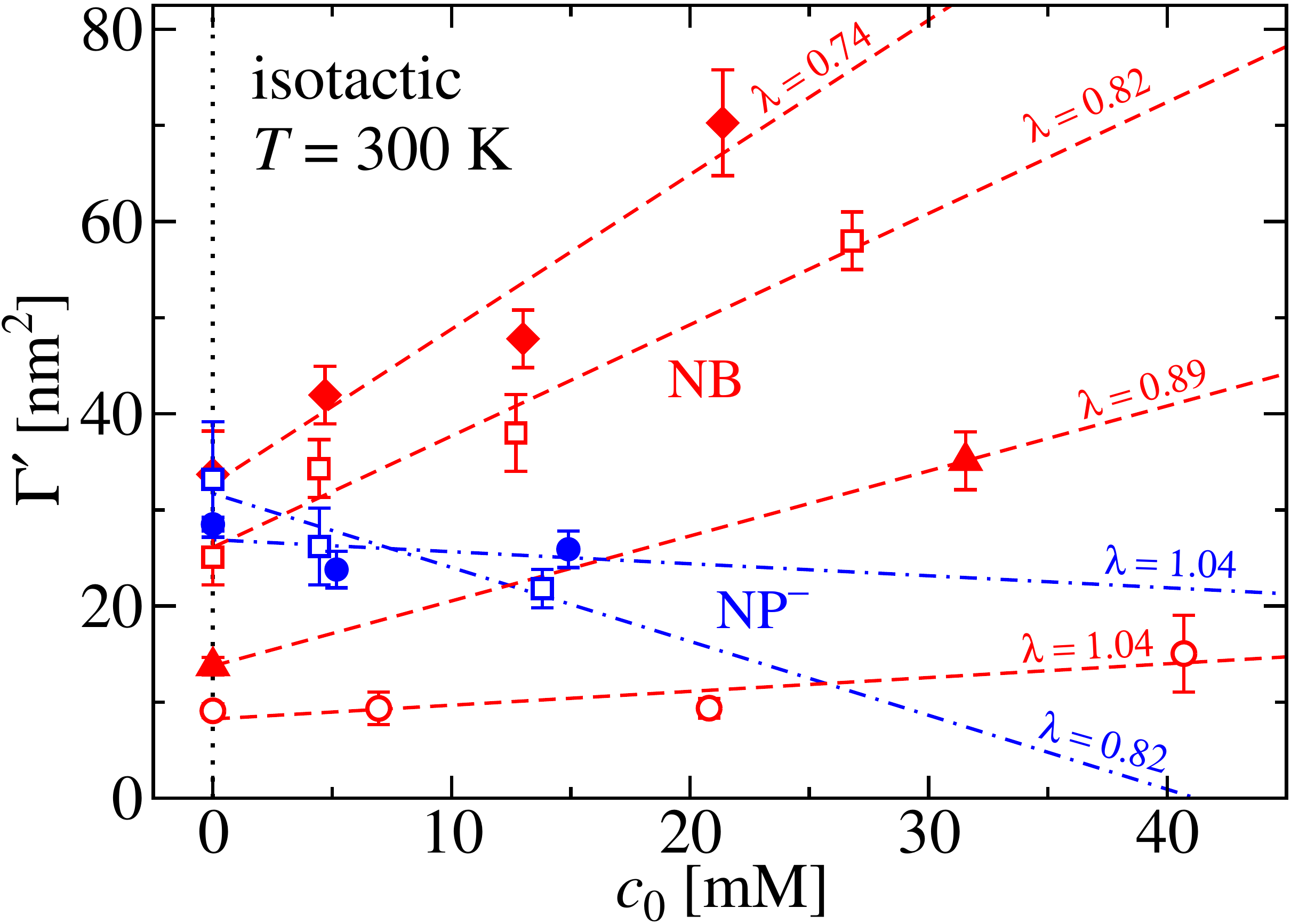}
\end{center}\end{minipage}
\caption{{Dependence of adsorption coefficients $\Gamma'$ on the concentration of NB~(red data points) and NP$^-$~(blue data points) for different elongations of the isotactic chain. The dashed lines represent linear fits to the data points.}}
\label{fig:Gamma_c0}
\end{center}\end{figure}
{
Figure \ref{fig:Gamma_c0} shows the adsorption coefficients $\Gamma'$ in dependence of the bulk concentration for NB and NP$^-$. 
As can be seen, for low to moderate concentrations, the adsorption coefficient increases approximately linearly with the bulk concentration $c_0$  for NB, thus we can write
\begin{equation}
\Gamma'=\Gamma_0'+\Gamma_1' c_0.
\label{eq:Gamma2}
\end{equation}
Here, $\Gamma_0'$ corresponds to the adsorption coefficient in the infinite dilution limit, $c_0\to 0$, and $\Gamma_1'$ captures the collective effects of the solutes.
The coefficient $\Gamma_1'$ depends significantly on the chain elongation $\lambda$: it becomes larger for more loose chains.
While the adsorption is cooperative for NB, it has the opposite effect in the case of NP$^-$.\\
Qualitative insights into the concentration effects can be gained from the virial expansion approach. In \SItext\ we derive \Eq~\ref{eq:Gamma2} for the adsorption of a weakly interacting gas on a flat surface on the level of the second virial coefficient. In that simple model, we obtain the relation
  $\Gamma_1'\propto -B_2^\trm{(2D)} \Gamma_0'^2$, where $B_2^\trm{(2D)}$ stands for the surface second virial coefficient. The latter one provides a measure for an effective interaction between	 two solute molecules. For an attractive interaction between solutes, as in the case for NB--NB, the surface virial coefficient is negative, $B_2^\trm{(2D)}$\,$<$\,0, and correspondingly $\Gamma_1'$\,$>$\,0. 
On the contrary, two NP$^-$ ions electrostatically repel, and thus $B_2^\trm{(2D)}$\,$>$\,0, whereby the exact value is governed by the amount of salt in the solution (note that for the salt-free case the second virial coefficient diverges).
  This means that as the first NP$^-$ adsorbs on the polymer, it becomes more difficult for the second NP$^-$ to adsorb.\\
Furthermore, the result of the virial expansion suggests the quadratic dependence of $\Gamma_1'$ on $\Gamma_0'$, that is, the higher the adsorption at low concentration, the larger is the cooperativity effect at higher concentrations. This trend can be qualitatively observed in \Fig~\ref{fig:Gamma_c0} where the curves with larger y-intercepts ($\Gamma_0'$) have larger slopes ($\Gamma_1'$).
\\
The above results imply that for NB and NP$^-$ the collective effects
 on adsorption become important in the mM range. These concentrations are, however, usually not reached in the catalytic experiments, where they are of the order of $c_0$\,$\sim$\,0.1 mM~\cite{dzubiellaAngew2012}, thus deeply in the infinite-dilution regime.
}

\subsection{Solute partitioning in a swollen PNIPAM hydrogel}
So far, we focused on adsorption details of different solutes on an isolated elongated PNIPAM chain at different conditions. 
The outcome of the simulations was expressed in terms of a {adsorption coefficient}  $\Gamma'$ defined by \Eq~(\ref{eq:Gamma1}).

\begin{figure}[t]\begin{center}
\begin{minipage}[b]{0.5\textwidth}\begin{center}
\includegraphics[width=\textwidth]{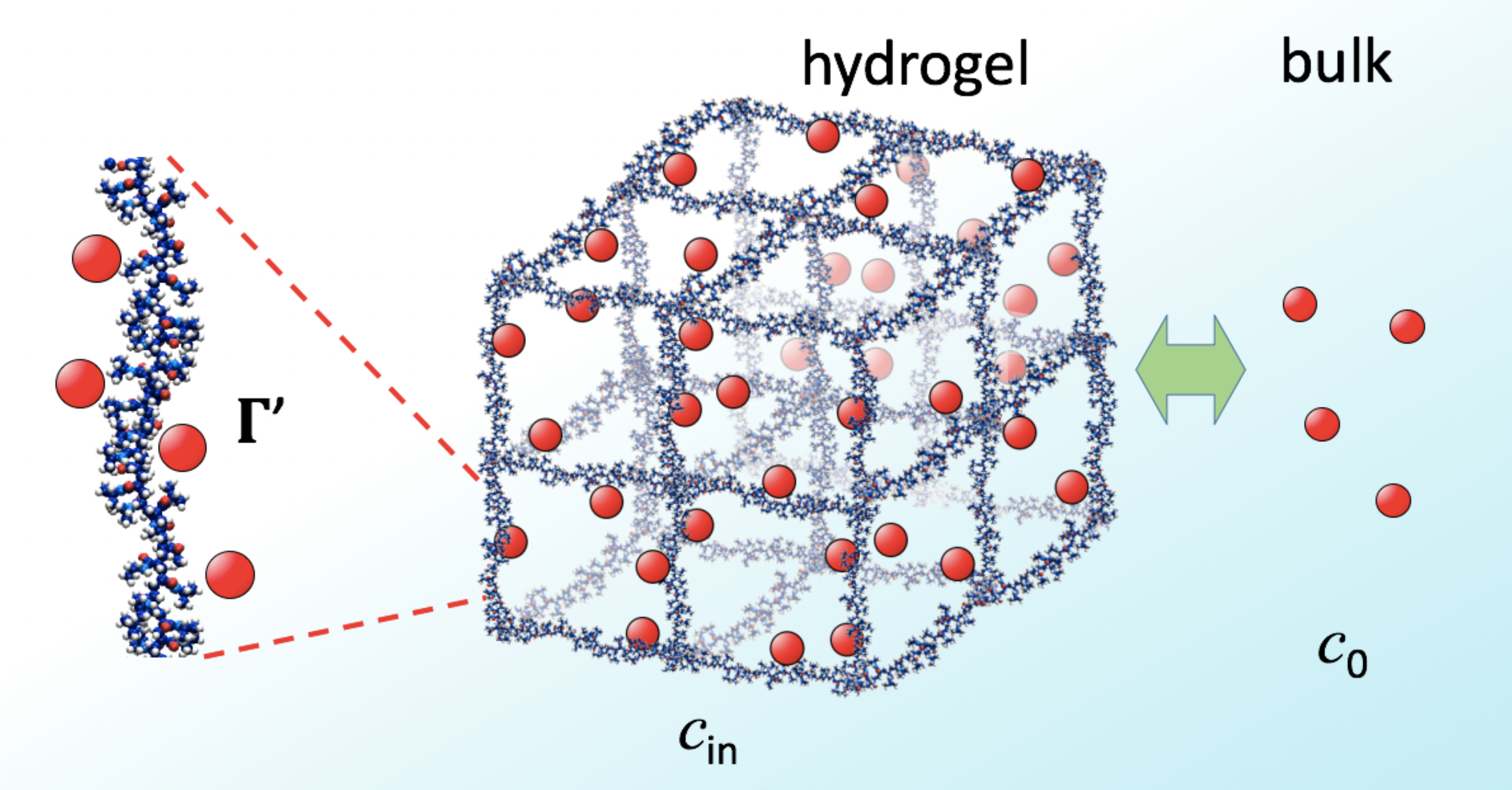}
\end{center}\end{minipage} 	
\caption{Schematic illustration of a swollen hydrogel model in contact with a bulk solution reservoir. The partitioning of the solute in the gel, $c_\trm{in}/c_0$, is estimated from the {adsorption coefficient} $\Gamma'$ on an isolated polymer chain.}
\label{fig:gel}
\end{center}\end{figure}
We now introduce a simple analytical model that relates the single-chain adsorption data, expressed in terms of {adsorption coefficient} $\Gamma'$, to the global 
solute partitioning and the solvation free energy of a solute from bulk solution into a swollen PNIPAM hydrogel environment.

For this, we consider a model of a very diluted hydrogel, where adjacent polymer chains are elongated and far apart and do not interfere with each other. We also neglect the influence of cross-linkers in the gel (suitable model also for a dilute brush). The gel is in chemical equilibrium with a bulk water solution with a solute concentration $c_0$, such that the gel and the bulk can exchange solute molecules, see \Fig~\ref{fig:gel}. 
The partitioning of the solute, that is, the ratio of the concentrations inside ($c_\trm{in}$) and outside ($c_0$) the gel is related to the solvation free energy $\Delta G$ (with respect to the bulk)  of the gel,
\begin{equation}
\frac{c_\trm{in}}{c_0}=\rme^{-\Delta G/\kB T}.
\label{eq:partition}
\end{equation}
Assuming that the gel is very dilute, we can estimate the concentration inside the gel from the data of adsorption on a single isolated chain, $\Gamma'$, as $c_\trm{in}={N_\trm{in}}/{V}=(c_0 V+\Gamma' c_0 \lambda L_\trm{c})/V$.
Here, the first term in the parenthesis, $c_0 V$, corresponds to the non-adsorbed solute background in the gel, whereas the second term is the number of the adsorbed solutes on the chain with a total contour length $L_\trm c$, see \Eqs~(\ref{eq:lambda}) and~(\ref{eq:GammaTOT}). The solute partitioning then expresses as
\begin{equation}
\frac{c_\trm{in}}{c_0}=1+\lambda\Gamma'\biggl(\frac{L_\trm{c}}{V}\biggr).
\end{equation}
The quantity $({L_\trm{c}}/{V})$ represents the total contour length of the polymer in the volume $V$. It can be expressed in a more standard way via the polymer volume fraction $\phi_\trm{p}=V_\trm p/V$. Considering the polymer as a cylinder with the radius $R_0$ (defined by the Gibbs dividing surface for the elongated chain, \Eq~(\ref{eq:R0})), its volume can be expressed as $V_\trm p=\pi R_0^2 L_\textrm c$, which yields
\begin{equation}
\biggl(\frac{L_\trm{c}}{V}\biggr)=\frac{\phi_\trm p}{\pi R_0^2}.
\label{eq:cinc0}
\end{equation}
The solvation free energy can now be written as
\begin{equation}
\Delta G=-\kB T\log\biggl(1+\lambda\Gamma'\frac{\phi_\trm p}{\pi R_0^2}\biggr).
\label{eq:G}
\end{equation}

In order to justify the model, we overview typical values of a PNIPAM hydrogel.
Typically, a PNIPAM hydrogel network is cross-linked by 5~mol~\% of crosslinkers with 4-fold connectivity~\cite{dzubiellaAngew2012}. That means, that an average polymer sequence between two crosslinkers  is 10 monomers long, which corresponds to the contour length of around $L_\trm c$\,$\approx$\,2.6~nm, as follows from our simulations. Since in the swollen state, the chains are almost fully extended, this is roughly also the distance between neighboring chains in the gel. 
As seen from the RDF in \Fig~\ref{fig:RDF-NB}b, we expect two neighboring chains to start interfering once their separation falls below $\sim$\,2~nm. With the estimated 2.6~nm of nearest-neighbor separation, we justify the single-polymer model used in our simulations.
The polymer volume fraction $\phi_\trm p$ sensibly depends on details of a hydrogel, temperature, and other external variables.
Undergoing the transition from the swollen to the collapsed state, the volume of the PNIPAM hydrogel with 5~mol~\% of crosslinkers shrinks by around a factor of 10~\cite{mashelkarMacromol1997, ravaineGlucose2006, dzubiellaAngew2012}.
 Since the water amount in the collapsed state is very low, we conclude that the water amount in the swollen state is roughly 10-times larger than the amount of the polymer. This means that the PNIPAM volume fraction in the swollen state is typically around $\phi_\trm{p}\!\sim$\,0.1.

Inserting these values in \Eq~(\ref{eq:G}) for NB, that is, $\Gamma'\!\approx$\,25~nm$^2$, $R_0\!\approx$\,0.5~nm, $\lambda\!\approx$\,0.8, and $\phi_\trm p\!\approx$\,0.1, we obtain the solvation free energy $\Delta G\!\approx$\,$-$1.3~$\kB T$ and partitioning, \Eq~(\ref{eq:partition}), $c_\trm{in}/c_0\!\sim$\,4.
The partitioning is an important quantity, \eg,  in the nanoreactor-based catalysis, because the total catalytic rate in a nanoreactor is proportional to the reactant concentration in the gel, $c_\trm{in}$. It is typically estimated by partitioning experiments~\cite{jiaJMCA2016}.
The current simulation results not only provide an alternative way to obtain such a quantity but provide insights on the local distribution and on the nature of the partitioning.

Note that the results based on \Eqs~(\ref{eq:cinc0}) and (\ref{eq:G}) are only a rough estimate, which considers many simplifications. The regions where two chains come close together may have an enhancing effect on the adsorption. Furthermore, crosslinkers could act as an important player. Nevertheless, the simple model provides principal insights into qualitative characteristics of the partitioning of molecules in swollen polymer architectures.

Finally, it is interesting to comment on the fact that the result \Eq~(\ref{eq:G}) depends on 
elongation $\lambda$. As a consequence this implies that 
partitioning has an implicit but strong temperature dependence in polymer architectures close to the volume transition where the chain extension 
can vary significantly. Furthermore, another implication is that external mechanical forces, such as longitudinal elastic stretching or compression can 
also lead to a controlled modification of partitioning of reactants in the polymer matrix.

\section{Conclusions}

In summary, we performed a detailed analysis of solute adsorption characteristics on a PNIPAM polymer chain in terms of atomistic explicit-solvent computer simulations. Employing a model of an elongated PNIPAM chain (mimicking a swollen polymer part), we explored representative classes of chemical compounds found  in small ligands and reactants.  We devoted special attention to adsorption properties of nitrobenzene and 4-nitrophenol, the key reactants in the benchmark catalytic experiments in thermoresponsive nanoreactors~\cite{songAdvMat2008, dzubiellaAngew2012, ballauffChemSocRev2012}. We found that all details such as elongation of the chain, its tacticity, as well as temperature play a significant role on the adsorption propensity. An interesting implication of the dependence on polymer elongation 
is that adsorption and partitioning should be controllable by external mechanical forces, \eg, induced by crowded media or an atomic force 
microscopy tip~\cite{vonklitzing2011}. 

Comparing binding affinities of various solutes in detail revealed, for instance, that the nitro group in an aromatic molecule significantly facilitates the adsorption on the polymer. On the other hand, hydroxylation plays a minor role, which contradicts previous beliefs that the adsorption is predominantly governed by the hydrophilic and hydrophobic nature of reactants. 
{We have mainly focused on the infinite dilution regime of the solute, where the adsorption is linearly proportional to its concentration. We showed that for higher solute concentrations, 
mutually attractive solutes, such as NB, exhibit cooperative adsorption, whereas mutually repulsive species NP$^-$ hinder further adsorption.}
At the end, we introduced a model to extrapolate the microscopic adsorption to macroscopic partitioning of the solutes in swollen polymer architectures, such as hydrogels. The latter are used as components
in responsive nanocarrier applications, where the partitioning of solutes is a key parameter to tune and control the desired material function. 

One of the main verdicts that emerged from our study regarding responsive model nanoreactors~\cite{dzubiellaAngew2012} is that 4-nitrophenol, be it in the protonated or deprotonated state, exhibits very similar adsorption as the more hydrophobic nitrobenzene molecule, which leads to similar partitionings of both reactants in a swollen PNIPAM hydrogel. The experimentally measured difference in the reaction rates below the polymer volume transition~\cite{dzubiellaAngew2012} is therefore expected to be governed by other effects, some of which may include interactions between neighboring chains, specific interactions with cross-linkers, charged impurities in the gel, or subtle chemical reactions at the catalytic surfaces. 
The switch of the reactions above the polymer volume transition might stem from very different partitioning of solutes in the collapsed state of the polymer matrix which, however, is currently not well characterized in structural terms.  The insights and conclusions of this study will serve as important guidelines for the forthcoming studies on partitioning and mobility in responsive polymer matrixes and carrier systems in general.

\section*{Acknowledgments}
The authors thank Yan Lu, Daniel Besold, and Matthias Ballauff for useful discussions on carrier systems and Robert Scholz for helpful consultations on partial-charge calculations.
We acknowledge funding from the ERC (European Research Council) within the Consolidator Grant with Project No.\ 646659-NANOREACTOR.
RC and JD thank the Deutsche Forschungsgemeinschaft (DFG) for financial support.
The simulations were performed with resources provided by the North-German Supercomputing Alliance~(HLRN).
\newline

 \setlength{\bibsep}{0pt}
\bibliography{literature}
\bibliographystyle{rsc}
\end{document}